# Interfacial properties of MoS$_2$ thin films grown on functional substrates

*Hafiz Sami Ur Rehman[1], Nunzia Coppola[1], Alice Galdi[1], Sandeep Kumar Chaluvadi[2], Shyni Punathum Chalil[2], Pasquale Orgiani[2,3], Sara Passuti[3], Regina Ciancio[3], Paolo Barone[4], Luigi Maritato[1] and Carmela Aruta[4,\**

[1]Dipartimento di Ingegneria Industriale-DIIN, Università Degli Studi di Salerno, Fisciano, Salerno 84084, Italy
[2]CNR-IOM, Strada Statale 14 Km 163.5, Basovizza, Trieste 34149, Italy
[3]Area Science Park, Padriciano 99, Trieste, Italy
[4]CNR-SPIN, Via del Fosso del Cavaliere 100, Roma 00133, Italy

*carmela.aruta@cnr.it

## Abstract

Interface chemistry and defect formation in MoS$_2$ thin films grown on single crystal substrates critically determine the electronic structure of MoS$_2$ and thus can strongly modify material functionality relevant for many applications, including electronics, optoelectronics, and energy-related catalysis. We investigate MoS$_2$ grown on three technologically relevant substrates, namely SrTiO$_3$(111), c-axis Al$_2$O$_3$(0001) and 6H-SiC(0001). Experimental investigations by temperature dependent resistivity, photoemission spectroscopy and scanning transmission electron microscopy with coupled energy dispersive spectroscopy, with the support of theoretical calculation by Density Functional Theory, allow the identification of the substrate induced specific defects and their correlation with the electronic properties. Ti interdiffusion in SrTiO$_3$/MoS$_2$ generates donor-like states near the Fermi level, leading to metallic transport. Al$_2$O$_3$/MoS$_2$ exhibits a high density of sulfur-related defects that introduce localized states and yield nearly temperature independent conductivity. SiC/MoS$_2$ exhibits significant interface disorder resulting in a semiconducting temperature dependent resistivity, yet deviating from the ideal bulk-like behavior. These results demonstrate how substrate choice governs defect formation and ultimately dominates the electronic behavior of MoS$_2$ thin films, making the control of film/substrate interactions essential for the engineering of new functional devices.

## 1. Introduction

Among the various two-dimensional transition-metal dichalcogenides (TMDCs), MoS$_2$ has gained increasing attention for its distinctive fundamental properties and technological relevance across a broad range of electronic, optoelectronic, quantum-technology, and energy-conversion applications. [1–3] Notably, MoS$_2$ undergoes a transition from an indirect-gap semiconductor in the bulk to a direct-gap semiconductor in the monolayer limit, a feature that further enhances its importance for optoelectronic and quantum-technology applications.[4,5]

Beyond electronics, $MoS_2$ is also a highly promising material for photocatalysis, electrocatalysis, and solar-driven chemical conversion, thanks to its abundant edge sites, tunable band structure, and strong light–matter interactions, which enable reactions such as hydrogen evolution, nitrogen reduction, and $CO_2$ conversion.[6,7]

The properties of $MoS_2$ are strongly influenced by defect distribution. When $MoS_2$ is grown as a thin films on specific substrates, its structural, electronic, and optical responses become sensitive to the interface with the substrate, as the interlayer van der Waals bonding within the material is intrinsically weak.[3,8] Defect engineering and interface control, especially on oxides and polar substrates, have been largely used to tune the functional properties. [3,6] Sulfur vacancies or ion interdiffusion from the substrate can create localized states within the bandgap, modify the charge density, and enhance catalytic activity by providing reactive sites.[9,10] Particularly relevant is the choice of substrate in emerging quantum-technology applications based on $MoS_2$. Substrate-induced defects can strongly influence valley coherence, exciton fine structure, and spin–valley coupling, which are key ingredients for quantum emitters, valleytronic devices, and coherent excitonic platforms.[11–14]

In the ultrathin limit, the interface becomes extremely relevant, making interface engineering a powerful means to modulate their formation and, possibly, giving rise to new states not present in bulk $MoS_2$.[3,15] Different interfaces between film and substrate can induce specific modifications in band alignment, electrostatic potentials, chemistry and surface symmetry mismatch, eventually leading to charge transfer process, improved carrier mobilities, and the formation of different types of defects. These phenomena have direct relevance for planar device architectures that integrate $MoS_2$ on insulating or semiconducting substrates.[3]

In this work, we investigate polycrystalline $MoS_2$ films grown by Pulsed Laser Deposition (PLD) on $SrTiO_3$(111) (STO), c-axis $Al_2O_3$(0001) and 6H-SiC (0001)substrates, which are commonly used for device fabrication in semiconductor and oxide-electronics. However, direct growth on functional semiconductor substrates is crucial for realizing high quality $MoS_2$ based heterostructures with clean interfaces and scalable device integration. Polycrystalline morphology appears regularly in $MoS_2$ thin films, grown by chemical (CVD) or physical (magnetron and PLD) vapour deposition techniques.[16–19] However, literature evidence suggests that granularity enables band-gap tuning through quantum confinement, which improves optoelectronics and photocatalytic performance. [17,20,21] In addition, it has been reported that phonon-boundary scattering reduces thermal conductivity in thermoelectric devices[22,23] and polycrystallinity facilitates defect-driven switching in memristive devices.[24,25] Despite domain boundaries, polycrystalline $MoS_2$ can show transport behavior comparable to single crystals and is suitable for large-area and flexible electronics.[26,27] For hydrogen evolution reaction (HER), polycrystallinity increase active sites and their efficacy.[28]

The STO, $Al_2O_3$ and SiC substrates used in this work are all wide-band-gap materials, thus providing excellent platforms for $MoS_2$ based electronic and optoelectronic devices. Their hexagonal surface symmetry exhibits a reasonable degree of compatibility with the hexagonal lattice of $MoS_2$, thereby making possible the growth of samples with good crystal properties. STO provides a high-K dielectric response which allow efficient charge carrier modulation in field-effect transistors.[29,30] Its polarizable lattice provides strong dielectric screening of Coulombic charge impurities. The formation of a covalent-like interaction between the sulfur atoms of $MoS_2$ can leads to charge transfer between the substrate and the $MoS_2$ layer.[31–33] $Al_2O_3$, with its chemical stability, is a common substrate for growing MoS2 films and is widely used as a high-k gate dielectric in $MoS_2$ transistors. Several reports have highlighted that $MoS_2$ samples show a better crystallinity in the case of $Al_2O_3$ with respect to $SiO_2$/Si substrate explained in terms of dipole–dipole interactions existing between S and Al atoms, which allow the better alignment of the $MoS_2$ layers.[34,35] In contrast, despite the promising potential of $MoS_2$ on SiC, the crystal quality achieved so far has not met the requirements of advanced electronic devices. A good crystallinity and performance for transistor and memristive applications was reported in case of $MoS_2$ grown by CVD.[36] Given the high thermal conductivity, large breakdown field, and excellent radiation stability of SiC, which make it particularly relevant in device applications, together with the theoretically demonstrated charge transfer at the interface, the SiC/$MoS_2$ interface provides significant potential for exploring new functionalities.[37]

We employ PLD to grow the $MoS_2$ films, as the versatility of this technique enables the deposition of layers with different compositions with high reproducibility.[38–40] Although less explored than other conventional growth methods of TMDC, PLD has shown strong potential for the controlled synthesis of high-quality $MoS_2$, enabling stoichiometric ultrathin films on a wide variety of substrates, tunable single-layer growth, and fine control of crystallinity, thickness, and morphology through plasma-energy regulation.[18,19,41] The flexibility and scalability of PLD allows engineering interfaces and heterostructures, and optimizing film quality for next-generation TMDC-based devices.

Our aim is to uncover how substrate interactions and defect profiles combine to modify the electronic landscape of $MoS_2$. To address these aspects, we combine X-ray photoemission spectroscopy (XPS) and high-resolution transmission electron microscopy (HRTEM) coupled with Energy-Dispersive Spectroscopy (STEM-EDS) together with Density Functional Theory (DFT) calculations, to identify the substrate-induced defects in $MoS_2$ grown on the three selected substrates and correlate them with the temperature dependent resistivity behaviour observed in thicker films. Our findings highlight how substrate selection governs defect formation and interfacial chemistry, ultimately shaping the electronic structure and transport properties of $MoS_2$ thin films. Understanding these mechanisms is a critical step toward the rational design of next-generation planar devices and catalytic systems that rely on few-layer $MoS_2$ with engineered optical, electronic, and chemical properties.[38,39]

## 2. Results and discussion

Temperature-dependent electrical transport measurements on $MoS_2$ films 60–80 nm thick, all grown under the same conditions but on different substrates, STO, $Al_2O_3$ and SiC, are reported in Figure 1. It can be observed that markedly different behaviours occur depending on the substrate. STO/$MoS_2$ exhibits good conductive behaviour, with a very low room-temperature resistivity of 0.22 mΩ·cm. In contrast, SiC/$MoS_2$ shows a more resistive semiconducting behaviour, with a slightly wavy temperature dependence and a higher room-temperature resistivity of 3.8 mΩ·cm. $Al_2O_3$/$MoS_2$ displays an intermediate trend, essentially semiconducting, but with the highest room-temperature resistivity among the three, reaching 7.3 mΩ·cm. The three temperature-dependent transport curves indicate that the electronic behaviour of MoS$_2$ is strongly influenced by the underlying substrate. The metallic-like response observed for STO/MoS$_2$ suggests that this interface may promote efficient charge transfer, possibly through favourable interfacial electronic coupling or substrate-induced modification of the electronic structure. The highly resistive, slightly wavy semiconducting behaviour of SiC/MoS$_2$ is indicative of interface-induced states that weakly modulate the activation behaviour. Al$_2$O$_3$/MoS$_2$ shows an intermediate, essentially semiconducting trend, which could arise from moderate interface disorder that limit the carrier mobility.

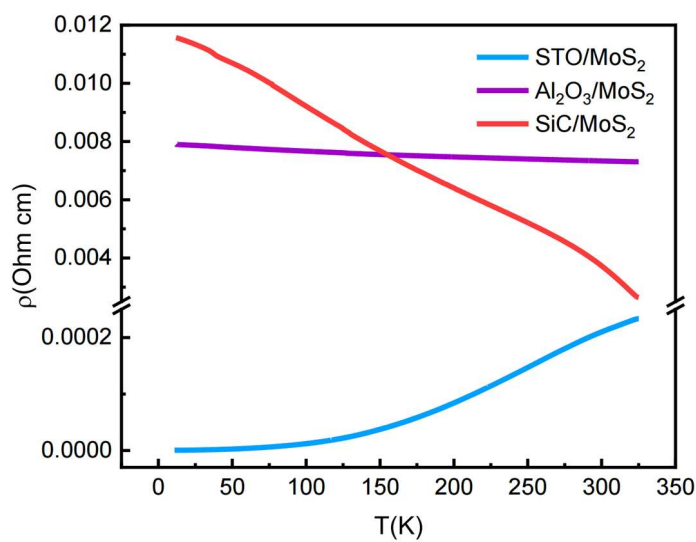

*Figure 1. Resistivity measurements as a function of temperature of MoS$_2$ films grown on STO, Al$_2$O$_3$ and SiC substrates.*

Overall, the distinct behaviours observed for the three substrates are consistent with substrate-dependent variations in interfacial strain, electronic coupling and defect distribution, which affect the bulk sample properties. Previous research on similar samples demonstrated that substrate-induced strain is not sufficient to explain the structural results obtained by X-ray diffraction (XRD).[40] The most probable in-plane mismatch between the MoS$_2$ film and the different substrates is approximately +0.6% for STO, +0.1% for Al$_2$O$_3$, and −2.8% for SiC. In contrast, the films exhibit a systematic increase of the c-axis lattice parameter with respect to bulk, following the trend SiC > STO > Al$_2$O$_3$, with the latter being the closest to the bulk 2H-MoS$_2$ value.[40] We can expect the presence of domains of different dimension, where the lattice mismatch is locally significant. However, the XRD results suggest that additional, more relevant effects occur at the interface and are likely responsible for the observed transport behaviour.

To clarify the origin of these interface related phenomena, we performed XPS measurements by synchrotron radiation on ultrathin films only 2 ML thick, a thickness specifically chosen to enhance interface response. The XPS experiments were carried out in situ by transferring the samples directly from the PLD to the analysis chamber without air exposure, a crucial aspect given the extreme sensitivity of 2 ML films to ambient contamination.

The XPS core level measurements of Mo 3d and S 2p are shown in Figure 2 a) and b), respectively. It can be observed that the shape of the Mo 3d signal does not change significantly across the $MoS_2$ films grown on the three different substrates, except for the peak broadening, which increases from STO to $Al_2O_3$ to SiC and thus indicates higher degree of disorder and the presence of additional components associated with defect states. More pronounced modifications can instead be observed in the profile of the S 2p core levels, although all of them exhibit a clear asymmetry towards the higher BE region. In the $STO/MoS_2$ interface, the spectrum is relatively sharp and well-defined. Moving to $Al_2O_3/MoS_2$, the peaks broaden slightly. In the $SiC/MoS_2$ sample, the broadening is most pronounced, indicating a higher degree of disorder or the presence of additional chemical environments. These variations reflect substrate-induced modifications in the sulfur chemical states and point to a progressive increase in defect density from STO to $Al_2O_3$ to SiC. We therefore performed a fit of S 2p spectra, to highlight the main differences in the shape of the S 2p core levels, shown in Figure 3. All spectra are fitted with two $S\ 2p_{3/2}$-$S\ 2p_{1/2}$ doublets. Doublet A corresponds to stoichiometric sulfur and appears at almost the same BE for all samples. Doublet B is left free to vary and contains the main differences among the three samples. More information about the fit are reported in Supporting Information.

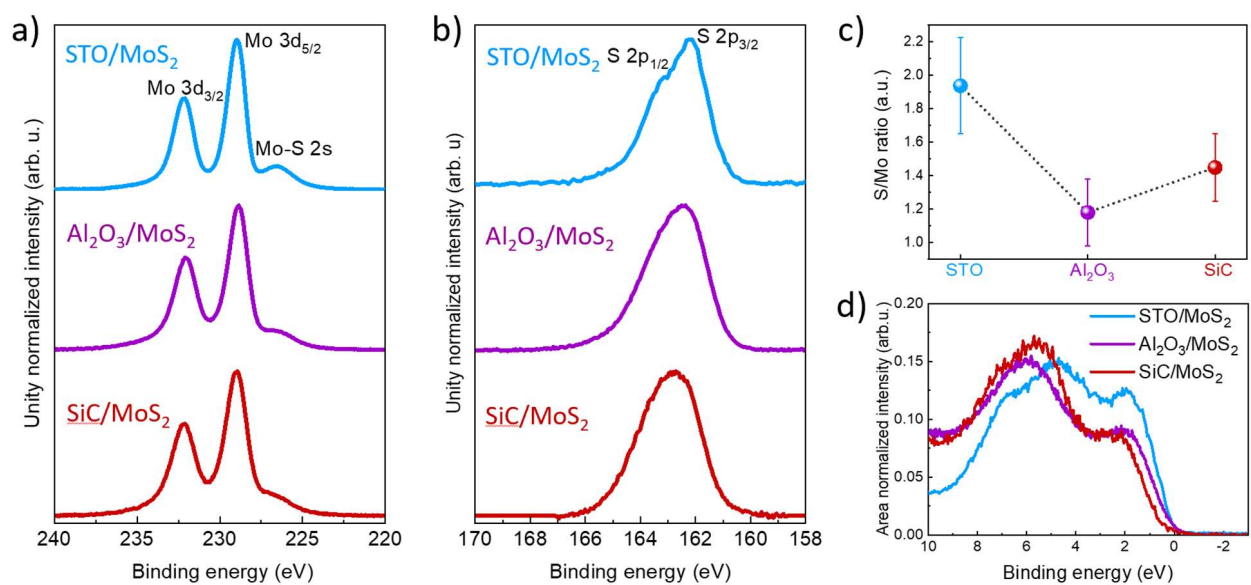

*Figure 2. XPS measurements on 2ML $MoS_2$ films grown on STO, $Al_2O_3$ and SiC. a) Mo3d core level experimental spectra together with multicomponent fit; b) S2p core level experimental spectra; c) Mo/S ratio obtained from a) and b) measurements; d) VB measurements.*

The calculation of the S/Mo ratio derived from the S 2p and Mo 3d core levels in the XPS spectra, is reported in Figure 2 c). All three samples exhibit sulfur vacancies, but the one grown on STO is the closest to stoichiometric 2 value, with S/Mo=1.94. The sample grown on $Al_2O_3$ shows the lowest ratio S/Mo=1.18, while the SiC/MoS$_2$ sample displays an intermediate value of S/Mo=1.45. The decreasing trend of the Mo–S 2s related feature in Figure 2 a) is consistent with the reduction of the S/Mo ratio.

In Figure 2 d) the VB measurements show that, compared to the MoS$_2$ on SiC, there is a increasing of Mo 3d states near the Fermi ($E_F$) level in the MoS$_2$ on Al$_2$O$_3$ sample and an even stronger depletion in the case of STO substrate. The calculated valence band maximum (VBM) values result +0.42 eV (SiC/MoS$_2$), +0.026 eV (Al$_2$O$_3$/MoS$_2$) and +0.02 eV (STO/MoS$_2$), as shown in Supporting Information. While the shift of the VBM closer to $E_F$ from SiC to STO is consistent with the better conduction observed for STO compared to SiC (Figure 1), the correlation between the VBM extending beyond $E_F$ and the transport properties of the sample grown on Al$_2$O$_3$ is less straightforward.

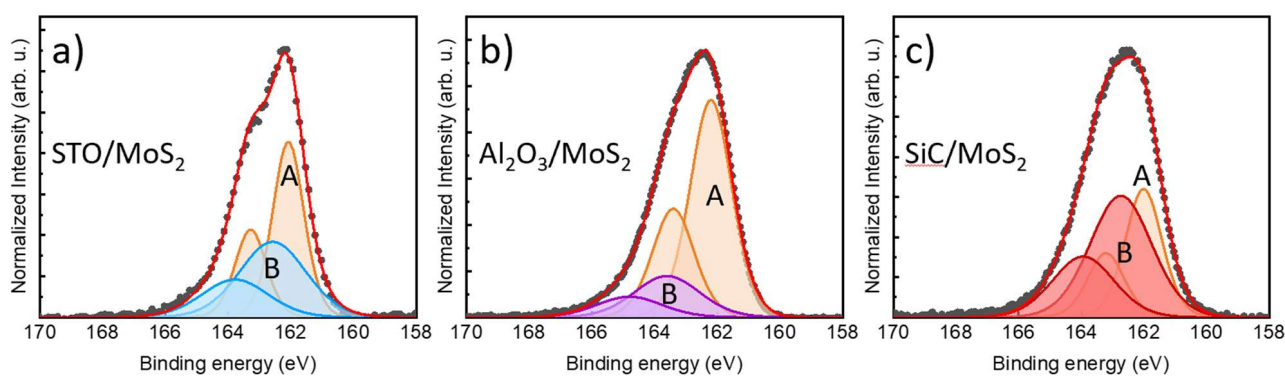

*Figure 3 Two components fit of 2Ml thick MoS$_2$ films on a) STO, b) Al$_2$O$^3$ and c) SiC substrates.*

To elucidate the structural origin of the defect-induced electronic states, STEM-EDS mapping was performed across the film/substrate interface to probe interfacial chemical disorder and elemental interdiffusion. Although the XPS results presented in this work refer to ultrathin films, the STEM investigations were carried out on thicker films grown under identical conditions. The use of thicker samples allowed the preparation of extended and more homogeneous cross-sectional regions during FIB-SEM specimen preparation, facilitating a more reliable structural and compositional analysis of the film/substrate interface. Cross-sectional lamellae were extracted by FIB-SEM to expose the film/substrate interface in cross-sectional geometry, allowing direct investigation of the interfacial region. STEM-EDS mapping was subsequently performed, acquiring elemental line scans across the interface together with spatially resolved chemical maps.

Concerning the sample deposited on STO, from the map overlays, we can observe a constant interdiffusion of Ti in the MoS$_2$ thin film across the length of the sample, of about 5 nm depth (Figure 4).

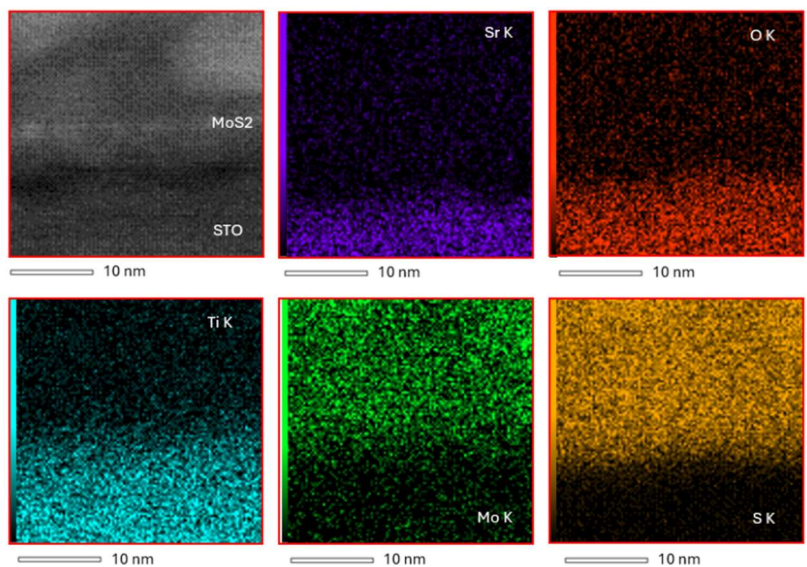

*Figure 4. Cross-sectional HAADF-STEM image of the MoS$_2$/STO interface (top left panel), together with corresponding STEM-EDS elemental maps acquired from the same region. The chemical maps display the spatial distribution of Sr K, O K, Ti K, Mo K, S K signals, respectively.*

The films deposited on Al$_2$O$_3$ (Figure 5) and SiC (Figure 6) result in being more stable. No interdiffusion is detected at the interface. However, a layer containing oxygen in the MoS$_2$ sample deposited on SiC of about 16 nm, which is observed across the whole film length, suggests an oxidation of the substrate surface prior to the film deposition.

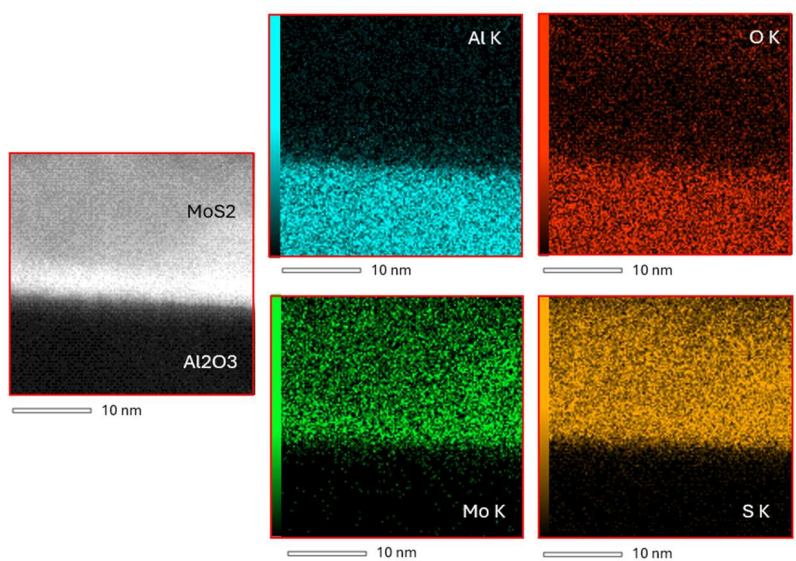

*Figure 5. Cross-sectional HAADF-STEM image of the MoS$_2$/Al$_2$O$_3$ interface (left panel), together with corresponding STEM-EDS elemental maps acquired from the same region (right panels). The chemical maps display the spatial distribution of Al K, O K, Mo K, and S K signals, respectively.*

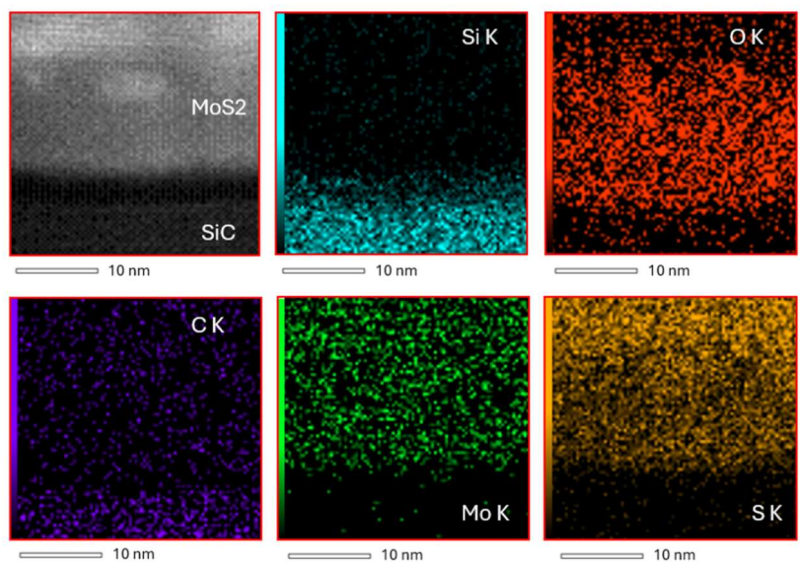

*Figure 6. Cross-sectional HAADF-STEM image of the MoS$_2$/SiC interface (top left panel), together with corresponding STEM-EDS elemental maps acquired from the same region. The chemical maps display the spatial distribution of Si K, O K, C K, Mo K, S K signals, respectively.*

DFT calculations were performed considering different types of defects.[42] The Density of States (DOS) and the corresponding energy shifts with respect to the core hole S$_0$ in the bulk-like sulfur ($E_{S0}$=-117.325 eV) were calculated. The results of all calculated defect shifts are reported in the Supporting Information. The effect of strain would result in very small shifts, falling well within the experimental uncertainty and the typical error associated with the multicomponent fitting of the XPS S 2p core level. Different sulfur defects are also considered, which are difficult to discriminate from STEM-EDS measurements. In addition, Ti and O substitution are calculated which can be relevant in case of STO/MoS$_2$ and SiC/MoS$_2$, as suggested by the STEM-EDS images. Based on these results, we try to identify the B component at higher binding energy in the fit of the S 2p spectra of Figure 3. Therefore, the defects associated with a positive chemical shift are the only considered for the interpretation of our results. In Figure 7 (a) the core-level energy shifts of selected defects, with respect to S$_0$ bulk-like reference, are indicated. In the S 2p fit the A component is associated to S$_0$. The comparison between the DFT shifts and the difference between B and A components, is highlighted using a histogram in Figure 7 (b).

In the case of STO/MoS$_2$ the Ti interdiffusion can be responsible for the observed B component in the XPS S 2p. The agreement between the DFT calculation and the XPS fit is quite good, as shown in Figure 7 (b). However, we cannot rule out the presence of additional sulfur-related defects. Since the sample grown on Al$_2$O$_3$ exhibits very sharp interfaces and there is no evidence from the STEM-EDS measurements of any Al or O interdiffusion from the substrate into the MoS$_2$, the B doublet can be first attributed to sulfur vacancies, as also indicated by the S/Mo ratio measured by XPS (Figure 2 c). These vacancies give rise to a shift of 0.048 eV, which is, however, very small and cannot by itself account for the B doublet. Therefore, we should consider the presence of additional sulfur-related defects. One possible type of defect can the S adatom at the hollow site $S^{ad}_{hol}$ which give rise to a DFT shift of 2.083 eV. This defect is accompanied by the shift of -0.057 related to the underlying S atom at the hollow site, which however is very small and not considered in our fit of Figure 3 (b). In this case the agreement between the DFT calculations and the XPS fit is less accurate. However, some differences between DFT and XPS are understandably expected, since DFT calculations have been performed on an ideal monolayer, while XPS measurements have been performed on an 2ML film grown on a single crystal substrates introducing interface effects. In the case of the SiC/MoS$_2$ sample, considering the oxygen diffusion observed in the STEM/EDS images (Figure 6), the oxygen defects substituting sulfur atoms produce higher energy shifts, but very small. Their shifts are 0.082 eV and 0.046 eV for the two sulfur atoms that are first neighbors to the oxygen defect. In contrast, oxygen adsorbed on top of a sulfur atom would induce a lower energy shift of -2.097 eV (see Supporting Information) which is not compatible with the spectral shape shown in Figure 3 (c). However, since the B doublet is very broad, oxygen substituted in the sulfur site cannot be the only relevant defect. Other defects, such as carbon substituting sulfur, which yields a shift of 0.144 eV, sulfur-related defects like $S^{ad}_{hol}$ or strain effects should be also considered, which must be very likely convoluted with a large amount of disorder, as evidenced by the very broad shape of the S 2p spectrum.

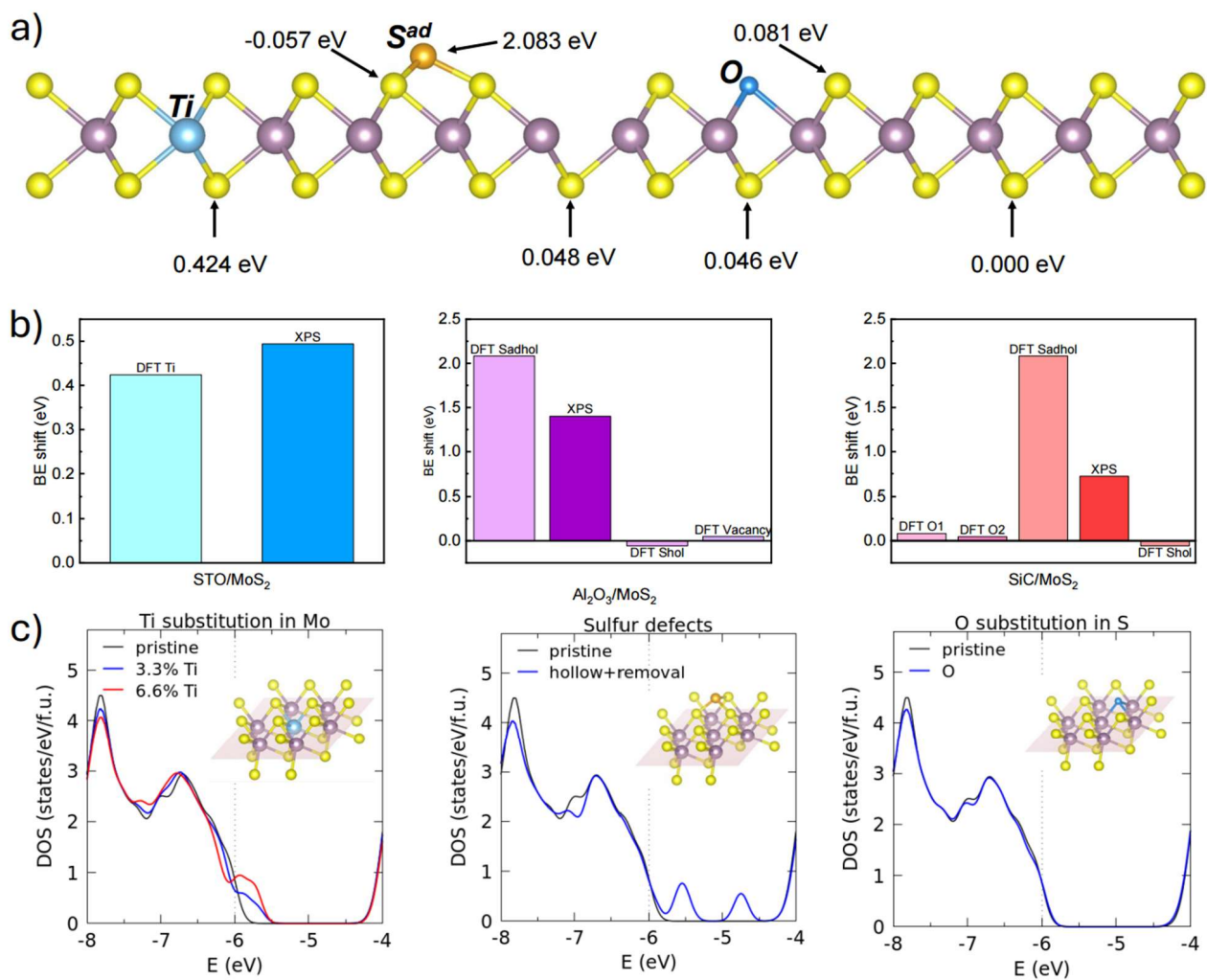

Figure 7. a) The core-level energy shifts of S 2p states with respect to the bulk-like sulfur used as a reference. b) Histogram of the XPS energy shifts of the B doublets used for the fit (in darker colour blue for STO, violet for Al$_2$O$_3$ and red for SiC subtrate), together with DFT results (in lighter colours) for different types of defects. c) DOS calculations for Ti interdiffusion (left panel), S in hollow position (center panel) and O interdiffusion (right panel).

The total DOS calculated by DFT shown in Figure 7 (c), when compared with the experimental VB spectra of Figure 2 (d) leads us to conclude that the most likely defect scenario is the Ti interdiffusion in STO/MoS$_2$, S adatoms at hollow sites in Al$_2$O$_3$/MoS$_2$ and O diffusion in a very disordered SiC/MoS2 system. This scenario is also in agreement with the transport measurements shown in Figure 1. Indeed, Ti substitution in STO/MoS$_2$ introduces defect states close to the VB edge, increasing the DOS at the Fermi level and shifting the VB edge upward, consistent with a p-type, degenerate electronic character that explains the metallic-like behavior. To the best of our knowledge, Ti substitution in MoS$_2$ or in other TMDCs has not been previously reported in the literature, nor the effect of Ti induced p-type doping. However, p-type doping in MoS$_2$ has been obtained by other substitutional transition metal dopants, as V, Nb or Ta, despite the intrinsic tendency of MoS$_2$ toward n-type transport driven by sulfur vacancies.[43–45] However, Ti cations in STO are quite mobile ad can diffuse through STO at the growth temperatures typically used for MoS$_2$. When Ti ions are migrated towards the surface in sulphur-rich environment, they are energetically favoured to form Ti–S bonds. The Ti incorporation into the MoS$_2$ lattice is facilitated by the close compatibility between the ionic radii of Ti and Mo, which allows Ti to substitute Mo in either octahedral or trigonal-prismatic coordination with S. In Al$_2$O$_3$/MoS$_2$, despite the high stability of the Al$_2$O$_3$ substrate surface, the formation of sulfur defects and vacancies can be related to the highly oxygen rich and strongly ionic surface of Al$_2$O$_3$, which can promote sulphur depletion from the interface forming volatile SO$_x$ species or O–S bonds, as well as the strong polarity can destabilize the interfacial MoS$_2$ layers. The sulfur defects and vacancies introduce localized states near the Fermi level shown in the calculated DOS of Figure 7 (c). These localized states hinder carrier mobility and suppress band-like transport, resulting in a resistivity that varies only weakly with temperature. In SiC/MoS$_2$, O substitution does not shift the Fermi level nor does it introduce localized defect states within the gap as indicated in the calculated DOS of Figure 7(c). We can conclude that the large amount of disorder can be responsible of the observed semiconducting behaviour. This interpretation is consistent with the XPS Mo 3d fits reported in the Supporting Information, which show a larger defective component for SiC compared to the other substrates. This can be explained by the chemically active nature of the SiC surface, containing native oxides, Si and C defects, all of which introduce reactive sites at the interface. Sulphur also has a strong chemical affinity for both Si and C, promoting parasitic reactions during growth. In addition, the lattice mismatch between MoS$_2$ and SiC is substantially large, leading to the structural disorder and the formation of misoriented domains. The combination of a highly reactive and structurally complex interface prevents the ordered growth of MoS$_2$ on SiC.

## 3. Conclusions

By examining the interplay of chemical environment and defect formation, both experimentally by XPS and STEM, and theoretically by DFT, we can interpret the temperature dependent behaviour of the resistivity of MoS$_2$ films grown on STO, Al$_2$O$_3$ and SiC substrates. For MoS$_2$ films grown on STO, the presence of substitutional Ti defects accounts for the resistivity behaviour, while in the cases of Al$_2$O$_3$ and SiC substrates, sulfur defects and chemical/structural disorder must be considered. As reported in the literature, Zhu et al. observed localized trap states arising from structural defects in MoS$_2$ films, which tend to reduce mobility.[46] However, according to a study by Kim et al. on monolayer MoS$_2$ grown under different sulfurization conditions, sulfur vacancies introduce shallow donor defects which, by donating electrons to the conduction band, screen the potential of structural defects and thereby enhance electron mobility.[47] In our case, sulfur vacancies are more abundant in films grown on Al$_2$O$_3$, whereas we assume that a high degree of structural and chemical disorder is present in the films grown on SiC. These findings are in agreement with the resistivity behaviour which increases strongly with decreasing temperature in the case of SiC, while the resistivity slope is smaller for the film on Al$_2$O$_3$, even though its room-temperature resistivity is higher than that of the SiC case.

The advanced and highly complementary techniques employed in this study enabled us to show that unexpected defects can form at the interface even with highly stable substrates. On one side, overlooking these interfacial defects may compromise the interpretation of experimental results, on the other side, understanding and controlling them open the possibility to tune interface properties. Considering the central role of the interface between film and substrate, not only for MoS$_2$ but also for other TMDCs and, more broadly, for technologically relevant thin film materials, our study provides insights into how interfacial phenomena govern the emergence of new electronic states which can be used to induce functionalities tailored to specific device applications.

## 4. Methods

**Sample deposition**

MoS$_2$ thin films were grown by PLD using a KrF excimer laser ($\lambda$ = 248 nm) at the NFFA APE beamline, Trieste.[48] The laser beam was focused onto a stoichiometric polycrystalline MoS$_2$ target (99.99% purity) with a fluence of approximately 2 J/cm$^2$. Depositions were carried out at a substrate temperature of 650 °C on three different single-crystal substrates: SrTiO$_3$ (111), c-plane Al$_2$O$_3$ (0001), and 6H-SiC (0001). The films were deposited under ultra-high vacuum (UHV) conditions, with a base pressure of ~ 10$^{-8}$ mbar, and the target-to-substrate distance was fixed at 5 cm. The growth rate was calibrated by X-ray reflectivity measurements and was determined to be ~ 200 laser pulses per unit cell, corresponding to one monolayer of MoS$_2$. After deposition, the samples were cooled to room temperature under the same pressure conditions as those used during growth.

**Transport measurements**

Electrical transport measurement were performed by four probe (Van der paw) method using Instron cryogenic low temperature system over the temperature range T=10K-325K. The ramp for the electrical transport is around 3K/min.

**Photoelectron spectroscopy**

XPS measurements were performed by synchrotron radiation after in-situ transferring from the PLD system directly connected to the transfer chamber of the APE-HE beamline at Elettra synchrotron in Trieste. The measurements were performed at room temperature using a hemispherical electrostatic analyzer with the sample at 45° with respect to the impinging linearly polarized light and normal to the surface. We collect the binding energies of the photoemission peaks were calibrated with respect to the Fermi level of Au reference samples.

**TEM**

HAADF-STEM imaging and STEM-EDS mapping were performed on a JEOL F200 TEM/STEM operated at 200 kV and equipped with a cold field-emission gun (cold FEG) and with a dual-detector EDS system. Imaging was carried out in STEM mode using a convergence semi-angle of 20 mrad, yielding a probe diameter of approximately 0.16 nm, while HAADF signals were collected over an angular range of 20–73 mrad.

Cross-section TEM lamellae were prepared from all three samples using a TESCAN Amber X plasma FIB-SEM via a site-specific lift-out procedure. The extracted lamellae were subsequently thinned and polished to electron transparency to enable STEM analysis of the film/substrate interface.

**Computational Details**

We performed first-principle calculations within the framework of Density Functional Theory (DFT) as implemented in the QUANTUM Espresso suite.[49] Norm-conserving pseudopotentials of the Trouiller-Martins type[50] have been adopted, while the Perdew–Burke–Ernzerhof (PBE) functional[51] has been used for the exchange-correlation energy within the generalized gradient approximation. We adopted a Gaussian smearing with a width of 0.01 Ry and a plane-wave cutoff of 100 Ry. Starting from bulk 2H-MoS2 with in-plane lattice parameter of 3.16 Å,[52] we constructed 15.80 x 16.42 Å supercells for simulating a single MoS2 layer, with 25 Å of vacuum separating periodic copies along the direction z perpendicular to the monolayer plane and adopting the truncation scheme for Coulomb interaction in the z direction.[53] A 3x3x1 mesh using the Monkhorst-Pack scheme[54] has been used for sampling the Brillouin zone, and the atomic positions have been optimized in all structures until forces were smaller than $10^{-3}$ Ry bohr$^{-1}$. Core-level binding energies of 2p states of S atoms have been calculated using the core-excited pseudopotential method.[55] To simulate photo-excited ions, we used the code atomic of the QUANTUM Espresso suite to generate a norm-conserving pseudopotential that includes a hole in the 2p subshell of sulfur. A scalar-relativistic approach has been adopted, implying that calculated core-level binding energies correspond to an average binding energy missing the splitting arising from spin-orbit coupling. The core-excited pseudopotential has been tested against known core-electron binding energies of simple molecules tabulated in Ref. [56], yielding an overall agreement with errors below 10%.

**Acknowledgements**

We acknowledge support from the Italian Ministry of Research under the PRIN 2022 Grant No 202228P42F with title "Transition metal dichalcogenide thin films for hydrogen generation" PE3 funded by PNRR Mission 4 Istruzione e Ricerca - Componente C2 - Investimento 1.1, Fondo per il Programma Nazionale di Ricerca e Progetti di Rilevante Interesse Nazionale PRIN 2022 – CUP B53D23003800006.
Research at SPIN-CNR was also supported by the project ECS00000024 "Ecosistemi dell'Innovazione"— Rome Technopole of the Italian Ministry of University and Research, public call n. 3277, PNRR— Mission 4, Component 2, Investment 1.5, financed by the European Union, Next Generation EU.
Maria Brollo and Jummi Laishram from Area Science Park are thankfully acknowledged for their support in lamellae preparation by FIB-SEM.
Angelo Bongiorno, from College of Staten Island, New York (USA) is also gratefully acknowledged for his insights and guidance on first-principles evaluation of chemical shifts of sulfur.

**Conflicts of Interest**

The authors declare no conflict of interest.

# Supporting Information

# Interfacial properties of MoS$_2$ thin films grown on functional substrates

*Hafiz Sami Ur Rehman, Nunzia Coppola, Alice Galdi, Sandeep Kumar Chaluvadi, Shyni Punathum Hafiz Sami Ur Rehman[1], Nunzia Coppola[1], Alice Galdi[1], Sandeep Kumar Chaluvadi[2], Shyni Punathum Chalil[2], Pasquale Orgiani[2,3], Sara Passuti[3], Regina Ciancio[3], Paolo Barone[4], Luigi Maritato[1] and Carmela Aruta[4,\**

[1]Dipartimento di Ingegneria Industriale-DIIN, Università Degli Studi di Salerno, Fisciano, Salerno 84084, Italy

[2]CNR-IOM, Strada Statale 14 Km 163.5, Basovizza, Trieste 34149, Italy

[3]Area Science Park, Padriciano 99, Trieste, Italy

[4]CNR-SPIN, Via del Fosso del Cavaliere 100, Roma 00133, Italy

*carmela.aruta@cnr.it

## X-ray photoemission spectroscopy fitting

We collect X-ray photoemission spectroscopy (XPS) survey scans, Mo 3d and S 2p core levels and valence band (VB) spectra on MoS$_2$ samples grown on SrTiO$_3$ (STO), Al$_2$O$_3$ and SiC substrates. Binding energy (BE) calibration is performed with reference to the adventitious C 1s at 284.8 eV. For the fitting procedure a Shirley function is assumed for background subtraction and a multicomponent deconvolution procedure is performed, using mixed Gaussian and Lorentzian line shapes. For the S 2p fitting of Figure XX of the main manuscript we used two components A and B made by doublets of the 2p3/2 and 2p1/2 spin-orbit splitting. We fixed as much as possible the fitting parameters. For all the doublets the splitting is fixed at 1.2 eV, the degeneracy ratio is 1:2 for the spin-orbit area ratio and the FWHM is a free parameter but the same for both peaks. The component A represents the stoichiometric sulfur.

The fit of the Mo3d shown in Fig. S1 was performed with two couples of 3d5/2 and 3d3/2 spin-orbit doublets for the Mo4+ and a higher binding energy (BE) contribution (defective Mo at higher valence state). A small contribution from the Mo6+ at even higher BE is hardly detectable, also owing to the in-situ transfer of the samples to the XPS analysis chamber, which limits the surface oxidation. In the fitting procedure the spin orbit splitting was fixed at 3.14 eV for all the doublets, the degeneracy ratio was 2:3 for the spin-orbit area ratio and the full width at half maximum (FWHM) was a free parameter but fixed at the same value for the two peaks of each doublet. At about 226.5 eV the Mo-S2s component is also included in the fit. In panel (a) the results for the MoS$_2$ films on different substrates (STO, Al$_2$O$_3$ and SiC) are reported, while in panel (b) the comparison between different thicknesses (1ML, 2ML and 3 ML) of the MoS$_2$ films grown on STO shows that the higher binding energy doublet is associated with the interface between film and substrate, as it tends to decrease as the thickness increases.

The valence band maximum (VBM) is obtained from the fitting of the VB spectra by the linear regression in the Mo 4d$z^2$ energy region (Fig. S2). The VBM is obtained from the intersection with zero intensity. [1]

In case of STO and Al$_2$O$_3$, the defect states calculated by DFT as described in the main text contribute to the shift of the Fermi level towards the valence band, thus decreasing the VBM.

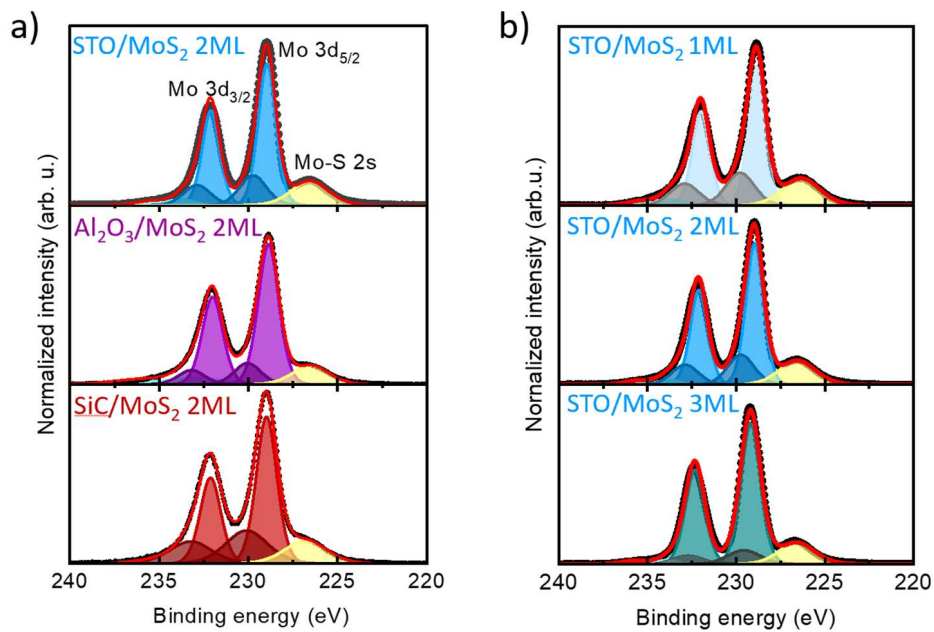

**Figure S1.** XPS Mo3d core level (a) of MoS$_2$ on STO, Al$_2$O$_3$ and SiC substrates, and (b) of MoS$_2$ on STO with 1ML, 2ML and 3ML thickness. The multicomponent fit is also shown.

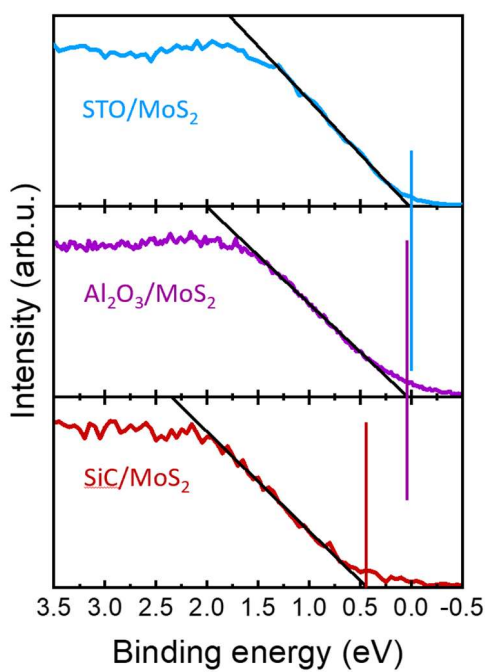

**Figure S2**. VB spectra close to the Fermi level to highlight the calculated valence band maximum (VBM): +0.42 eV (SiC/MoS$_2$), +0.026 eV (Al$_2$O$_3$/MoS$_2$) and +0.02 eV (STO/MoS$_2$).

The stoichiometry of the samples (Fig. XX of the main manuscript) is obtained from the $S^{2-}/Mo^{4+}$ total area ratio with relative sensitivity factors of 9.5 and 1.67 for Mo 3d and S 2p core levels respectively.

## Density Functional Theory calculations

Density Functional Theory (DFT) calculations of the S 2p shift for different type of defects are calculated as described in the main text and reported in Table XX with respect to sulfur in the bulk-like $MoS_2$.

| Chemical shifts (eV) of sulfur defects | $S_{rem}$ | $S_{api}$ | $S_{hol}$ | $S_{api}^{ad}$ | $S_{hol}^{ad}$ |
|---|---|---|---|---|---|
| | 0.048 | -1.507 | -0.057 | 0.870 | 2.083 |

**Table S1.** Calculated core-level energy shifts of sulfur defects with respect to ordinary S in a pristine $MoS_2$ monolayer. Shifts are calculated for sulfur atoms that are closest to the defect labeled by the subscript. Three kind of defects have been considered: a sulfur vacancy, a sulfur adsorbed on top of another sulfur and a sulfur adsorbed on the hollow site above the hexagonal $MoS_2$ lattice. Chemical shifts of adsorbed S defects, denoted by the superscript "ad", are also listed.

| strain | 0% | +2% | -2% |
|---|---|---|---|
| $S_{rem}$ | 0.048 | 0.026 | 0.074 |
| $S_{api}$ | -1.507 | -1.527 | -1.488 |
| $S^{ad}_{api}$ | 0.870 | 0.919 | 0.813 |

**Table S2.** Calculated core-level energy shifts with different strain conditions for the sulfur adsorbed on top of another sulfur and a sulfur adsorbed on the hollow site above the hexagonal $MoS_2$ lattice, already reported in Table S1.

| S chemical shifts (eV) due to other defects | Ti | Sr | Al | C | O |
|---|---|---|---|---|---|
| | 0.424 | 0.766 | 0.596 | 0.091/0.045 | 0.081/0.046 |

**Table S3.** Calculated core-level energy shifts of S 2p close to substitutional sites. We considered both substitution of Mo atoms with Ti, Sr, Al and of sulfur atoms with C or O. In the latter case, we report two values, corresponding to a hole in a nearest-neighbor sulfur atom belonging respectively to the same S layer where a sulfur atom has been replaced or to the other S layer.

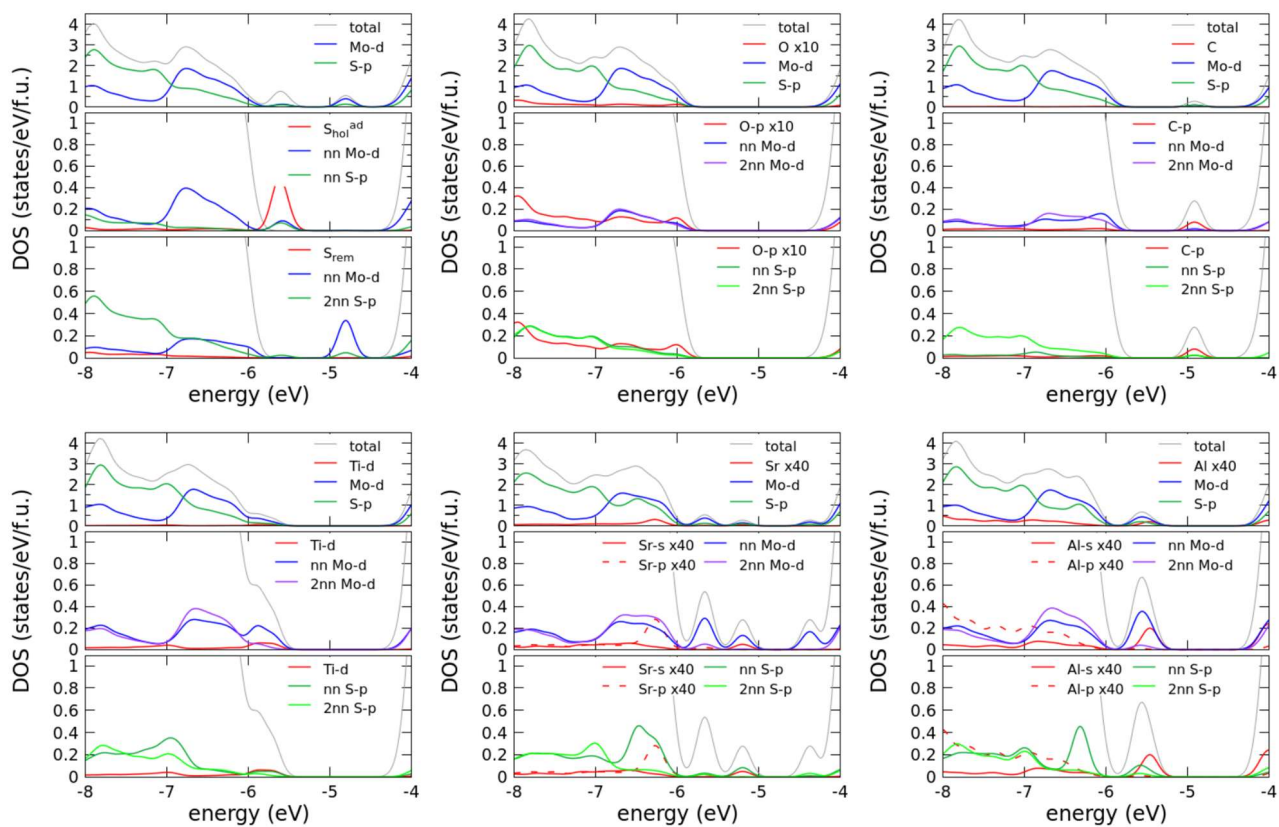

**Figure S3.** Project density of states (energy expressed with respect to the vacuum level) of the two most relevant sulfur defects used for the interpretation of our data (a sulfur vacancy and a sulfur adsorbed on top of another sulfur) and all the defects reported in Table S3.

**References**

[1] Lin, Y.-K. et al. Thickness-Dependent Binding Energy Shift in Few-Layer MoS2 Grown by Chemical Vapor Deposition. ACS applied materials & interfaces 8, 22637-22646 (2016).